## Generative artificial intelligence prompting strategies among Chinese university students across academic tasks

Lei Fan[1,*] and Fangxue Liu[1,2]
[1] Department of Civil Engineering, Xi'an Jiaotong-Liverpool University, Suzhou, 215123, China
[2] Department of Computer Science, University of Liverpool, Liverpool, L69 3BX, UK
*Corresponding Author (Email: Lei.Fan@xjtlu.edu.cn)

**Abstract**: Generative artificial intelligence (GenAI) is increasingly embedded in university learning, yet effective human–GenAI interaction depends not only on access to these tools but also on students' ability to formulate, evaluate, and refine prompts for academic tasks. This study examined university students' perceived GenAI prompting capability and reported prompting strategies. Questionnaire data from 234 students were analyzed using descriptive statistics, measurement modelling, scenario comparisons, structural associations, and qualitative coding of optional comments. Students reported using GenAI across diverse academic activities and generally expressed confidence in their prompting capabilities, particularly in providing relevant context and refining prompts following initial outputs. However, prompting strategies showed limited sensitivity to task types: specifying output format was more strongly endorsed in a creative workshop preparation scenario than in a research-oriented assignment, whereas other matched strategies showed no adjusted differences. Students also commonly endorsed both concise instructions and providing all relevant information at the outset, suggesting that effective brevity may depend more on selecting and organizing task-relevant information than on prompt length alone. By contrast, fixed prompt templates received lower endorsement than other prompting strategies. In the adopted factor model, perceived cognition showed the clearest unique association with perceived prompting capability, while open-ended responses highlighted prompt–intent alignment as a common challenge.

### 1. Introduction

Generative artificial intelligence (GenAI) has made support for explanation, writing, information seeking, and idea development readily available to university students. However, an accessible conversational interface does not eliminate the intellectual demands of effective human–GenAI interaction. Students must formulate appropriate requests, provide relevant context, and judge whether generated responses are accurate, useful, and suitable for the task. Previous surveys documented perceived academic benefits alongside concerns about accuracy, reliability, and responsible use (Chan & Hu, 2023; Fan, Deng, & Liu, 2025; Fan & Liu, 2026a). How students manage these interactions is therefore an educational issue rather than merely a question of technology adoption.

Prompt engineering refers to the deliberate formulation and iterative refinement of instructions, context, and constraints to guide GenAI toward a task-appropriate response. It includes decisions about defining a task, supplying relevant background or examples, specifying output requirements, evaluating generated responses, and revising prompts when necessary. Prompt patterns provide reusable structures for organizing such interactions (White et al., 2023). In educational settings, however, the value of prompting lies not simply in producing a better response but in whether the interaction supports the purpose of the academic task and enables critical evaluation of the resulting output. Research has linked aspects of prompt quality and artificial intelligence (AI) literacy to the quality of GenAI interactions, but improved generated output should not itself be interpreted as evidence of improved learning (Knoth et al., 2024; Fan, Tang, et al., 2025).

Important limitations remain in the literature. First, much of the survey research on student GenAI use has clarified patterns of adoption, perceived benefits, and concerns but has provided less detail about how students formulate requests during GenAI interaction. Second, empirical studies of prompting have often examined particular tasks or disciplinary settings, including introductory programming and undergraduate history (Nguyen et al., 2024; Woo et al., 2026). Such work has provided valuable evidence about prompting within specific learning activities, but broader evidence is still needed on how students report adapting prompting strategies across contrasting academic tasks.

To address these limitations, this study used a questionnaire to examine students' perceived prompting capability, strategies and difficulties. The study also investigated how four broader AI-related orientations (i.e., affect, cognition, behavioral intention, and ethics), were associated with perceived prompting capability and scenario-specific prompting strategies. Drawing on multidimensional accounts of AI literacy, these orientations were treated as related but distinct aspects of students' engagement with GenAI rather than as interchangeable indicators of prompting capability (Long & Magerko, 2020; Ng et al., 2021; Carolus et al., 2023).

Three research questions guided the analysis. RQ1: What patterns characterize students' perceived prompting capability, and reported prompting difficulties? RQ2: How do students' prompting strategies differ between the research-oriented assignment and creative workshop preparation scenarios? RQ3: How are affect, perceived cognition, behavioral intention, and ethics associated with perceived prompting capability and scenario-specific prompting strategies, and how is perceived prompting capability associated with strategy use across the two scenarios?

## 2 Literature review

### 2.1 Prompt engineering in higher education

In higher education, prompt engineering can be understood as the deliberate specification and iterative refinement of a task for a GenAI system. A prompt may define the problem, provide relevant context or examples, constrain the expected response, or request subsequent revision. White et al. (2023) organized such decisions into reusable prompt patterns, while Lo's (2023) CLEAR framework framed effective prompting in terms of being concise, logical, explicit, adaptive, and reflective within an information-literacy context. These frameworks provide useful conceptual vocabularies for describing how users structure interactions with GenAI, but they do not by themselves establish that applying a particular pattern improves learning. Their educational relevance therefore depends on how prompting is aligned with the task and how generated responses are subsequently interpreted, evaluated, and used.

Empirical research indicated that effective prompting involved more than entering an initial instruction. In a study of beginning programmers, Nguyen et al. (2024) found difficulties in articulating intentions, evaluating generated code, and revising prompts when outputs were inadequate. Zamfirescu-Pereira et al. (2023) similarly found that non-AI experts often relied on exploratory and trial-and-error prompting rather than systematically diagnosing why an interaction had failed. Together, these findings identified formulation, contextualization, evaluation, and revision as meaningful components of human–GenAI interaction. They also supported examining how students perceived their capability in these areas and which strategies they reported using, while recognizing that self-reported practices do not establish whether those strategies were implemented effectively.

Research further suggested that prompting practices were shaped by task and disciplinary context. Aruleba et al. (2025) examined computer-science students' prompting strategies and reflections across programming activities, whereas Woo et al. (2026) investigated changes in GenAI knowledge, self-efficacy, prompt artifacts, and reflections following a prompt-engineering workshop with undergraduate history students. Knoth et al. (2024) likewise examined relationships among AI literacy, prompt-engineering quality, and generated outputs across different tasks. Collectively, this work supported viewing prompting as a context-sensitive practice rather than as a single general technique. At the same time, the concentration of individual studies on particular tasks or disciplinary settings left scope for broader evidence on how students reported adapting strategies across contrasting academic and creative contexts.

## 2.2 Prompting and student learning

The relationship between prompting and learning requires a distinction between the quality of a GenAI-assisted product and the learner's knowledge, transfer, and evaluative judgment. Numerous empirical studies found that students valued GenAI for activities including writing, research, and idea development while also expressing concerns about reliability and ethical use (Chan & Hu, 2023; Fan & Liu, 2026b). Among engineering students, Fan, Deng & Liu (2025) reported strong perceived improvements in learning efficiency, while many respondents perceived little corresponding change in academic performance. These findings helped explain why students considered GenAI useful, but they neither identified which prompting practices produced the perceived benefits nor established learning effects.

Experimental evidence reinforced this distinction. In a randomized study of university students, Fan, Tang, et al. (2025) found that the ChatGPT-supported group achieved greater improvement in essay scores, whereas differences in knowledge gain and transfer were not statistically significant. Improved performance on a GenAI-assisted work therefore should not automatically be interpreted as improved learning. For prompting pedagogy, this distinction suggests value in considering not only the quality of the generated product but also whether students can explain their decisions, evaluate and verify generated information, and justify revisions. These are pedagogical implications of the existing evidence rather than outcomes tested directly in the present survey.

Self-regulated learning provides a useful interpretive lens because its models connect planning, monitoring, control, and reflection (Zimmerman, 2002; Panadero, 2017). Prompt formulation can represent planning, evaluation of a generated response can contribute to monitoring, and prompt revision can function as a control response when an output does not meet the task goal. However, reported revision alone does not demonstrate effective monitoring or self-regulation: students may repeatedly modify prompts without identifying why a response is inadequate. Constructive alignment provides a complementary pedagogical perspective by emphasizing consistency among intended learning outcomes, learning activities, and assessment (Biggs, 1996). Applied to prompting instruction, this perspective suggests that prompt formulation and output evaluation should remain connected to the disciplinary objectives that the academic task is intended to develop.

## 2.3 Contextual strategy use

Prompting is inherently context sensitive because the relevance of providing background, assigning a role, specifying an output format, imposing constraints, or revising an instruction varies with the task goal, information requirements, and expected product. The same strategy may therefore be applicable across contexts while serving different functions or carrying different importance in each. Prompt-pattern research has shown how reusable prompting structures can be adapted across problems and domains, while disciplinary studies have demonstrated that prompting practices are embedded in the requirements of particular learning activities (White et al., 2023; Aruleba et al., 2025; Woo et al., 2026). This literature supports examining strategy use within clearly specified contexts while reserving claims about strategy effectiveness for research that directly observes prompts, generated outputs, and relevant learning or performance outcomes.

# 3 Methodology

## 3.1 Study design and participants

The study used a cross-sectional online questionnaire administered via Wenjuanxing from early July to early September 2026. Convenience sampling was supplemented by purposive recruitment to improve representation across gender, degree level, academic discipline, and geographic regions of China. The study protocol was approved by the University Research Ethics Review Panel of Xi'an Jiaotong-Liverpool University on 8 Jun 2026. The survey was conducted in accordance with applicable ethical guidelines and regulations, and informed consent was obtained from all participants prior to their participation.

Disciplines were grouped into science, technology, engineering, and mathematics (STEM) and humanities and social sciences (HSS). Responses with identical numerical ratings across all rating items or indicating "none tried" in the prompting-method question were excluded, yielding a final analytical sample of 234 respondents. Participant characteristics are summarized in Table 1.

Table 1. Participant characteristics

| Characteristic | Category | Percentage (%) |
|---|---|---|
| Gender | Male | 43.2 |
| | Female | 56.8 |
| Degree level | Bachelor | 47.0 |
| | Postgraduate | 24.4 |
| | Three-year college diploma | 28.6 |
| Subject category | STEM | 51.7 |
| | HSS | 48.3 |

### 3.2 Questionnaire and measures

The questionnaire comprised three components: participant background information, GenAI use characteristics, and rating items assessing GenAI-related orientations and prompting strategies. Background and use questions covered frequency of GenAI use, commonly used tools, academic tasks for which GenAI had been used, and prompting methods previously attempted.

The main rating component included 27 items across seven domains. The first four domains contained three items each and assessed affective orientation (Domain 1; D1), perceived cognition (Domain 2; D2), behavioral intention (Domain 3; D3), and ethical orientation (Domain 4; D4). Domain 5 (D5) comprised five items assessing perceived prompting capability. Domains 6 and 7 assessed prompting strategies in two task scenarios, with five items each. Domain 6 (D6) focused on a research-oriented assignment examining the impact of GenAI on university students' employment, whereas Domain 7 (D7) focused on developing an appealing title and slogan for a workshop on GenAI and the future of education.

Four additional rating items assessed general prompting strategies: using brief instructions, structuring longer prompts, providing detailed information at the outset, and using a fixed prompt template. These items were analyzed separately because they reflected distinct interaction preferences rather than a common latent construct.

All 31 rating items used a five-point Likert-type scale ranging from strongly disagree (1) to strongly agree (5). All closed-ended questions were mandatory. The questionnaire concluded with an optional open-ended question asking respondents to describe difficulties they had experienced when prompting GenAI. The full questionnaire is provided in Appendix A.

### 3.3 Statistical analysis

Descriptive analyses included percentages, means ($M$), and item-level response distributions. For agreement-based items, endorsement was defined as selecting agree (4) or strongly agree (5). Internal consistency for each of the seven questionnaire domains was assessed using Cronbach's alpha ($\alpha$), with 95% confidence intervals based on 5,000 participant-level bootstrap resamples. The Kaiser–Meyer–Olkin (KMO) measure and Bartlett's test of sphericity (Kaiser & Rice, 1974; Tobias & Carlson, 1969) were used to assess the suitability of the correlation structure for factor analysis. These reliability and factorability diagnostics were treated as supporting evidence rather than as direct evidence of construct validity.

Contextual differences between the research-oriented assignment and creative workshop preparation scenarios were examined using paired-samples $t$ tests for four conceptually matched strategy items. For each comparison, we reported the creative-minus-research mean difference, its unadjusted 95% confidence interval, and the standardized paired effect size *dz*, calculated as the mean paired difference divided by the standard deviation of the paired differences. The third item in each scenario was excluded because the two items represented substantively different strategies. The Benjamini–Hochberg (BH) procedure was applied across the four tests to control the false discovery rate (Benjamini & Hochberg, 1995).

The measurement structure was examined using a seven-correlated-factor model in which each of the 27 domain indicators loaded only on its predefined factor, with no cross-loadings or correlated indicator residuals. A strict one-factor model was fitted for comparison. Both models used normal-theory maximum likelihood estimation based on the observed covariance matrix, treating the five-category indicators as approximately continuous (Rhemtulla et al., 2012). Model fit was evaluated using the chi-square statistic ($\chi^2$), comparative fit index (CFI), Tucker–Lewis index (TLI), root mean square error of approximation (RMSEA) with 90% confidence intervals, standardized root mean square residual (SRMR), standardized factor loadings, and latent-factor correlations. Accordingly, the reported fit indices were conventional maximum-likelihood rather than robust scaled ordinal statistics.

A structural model was then used to examine associations among the seven domains. Perceived prompting capability was regressed on the four GenAI-related orientation domains, while each scenario-specific strategy domain was regressed on the four orientations and perceived prompting capability. This specification yielded 14 structural paths, treated as a separate BH testing family from the four scenario comparisons. Standardized path coefficients were estimated using sandwich standard errors and delta-method confidence intervals (Oehlert, 1992). Because this structural specification was covariance-equivalent to the seven-factor measurement model, overall model fit could not establish the direction of the structural relationships and was not interpreted causally. Complete path estimates are reported in Section 4.8.

Responses to the optional open-ended question were analyzed using descriptive content coding, with multiple codes permitted for each response where appropriate. Percentages were calculated only among respondents who provided a comment and were therefore not interpreted as prevalence estimates for the full sample.

## 4 Results and interpretation

### 4.1 Internal consistency and factorability

As shown in Table 2, all seven domains demonstrated high internal consistency, with Cronbach's alpha ranging from 0.858 to 0.935. The lower bounds of all bootstrap 95% confidence intervals exceeded 0.80, indicating consistently strong reliability. Domain-specific KMO values ranged from 0.731 to 0.894, and the overall KMO for the 27 domain items was 0.964. Bartlett's test of sphericity was significant, $\chi^2(351) = 5413.80$, $p < 0.001$, indicating that the correlation matrix differed substantially from an identity matrix. Collectively, these results showed that the item correlations were suitable for factor analysis and that the seven domains had strong internal consistency. These diagnostics supported the subsequent measurement analyses but were not treated as evidence of construct validity in themselves.

Table 2 Domain summaries and internal consistency

| **Domain** | **Items** | **$\alpha$ [95% Confidence Interval]** | **KMO** |
|---|---|---|---|
| D1: Affective orientation | 3 | 0.908 [0.873, 0.934] | 0.753 |
| D2: Perceived cognition | 3 | 0.866 [0.823, 0.899] | 0.731 |
| D3: Behavioral intention | 3 | 0.861 [0.822, 0.894] | 0.732 |
| D4: Ethical orientation | 3 | 0.858 [0.806, 0.898] | 0.733 |
| D5: Perceived prompting capability | 5 | 0.935 [0.916, 0.949] | 0.894 |
| D6: Research-oriented assignment strategies | 5 | 0.903 [0.872, 0.926] | 0.836 |
| D7: Creative workshop strategies | 5 | 0.890 [0.853, 0.917] | 0.878 |

### 4.2 Academic uses of GenAI

As shown in Figure 1, students used GenAI across a range of academic tasks. The most frequently reported uses were report or paper writing (60.7%), literature searching and reading (59.4%), and report or paper polishing (57.7%). Learning new knowledge or explaining concepts (45.7%), data analysis (42.7%), and translation (40.2%) were also commonly reported. The high proportions for both writing and polishing suggest that students used GenAI extensively for both text generation and revision. Similar patterns have been reported in previous studies (e.g., Chan & Hu, 2023).

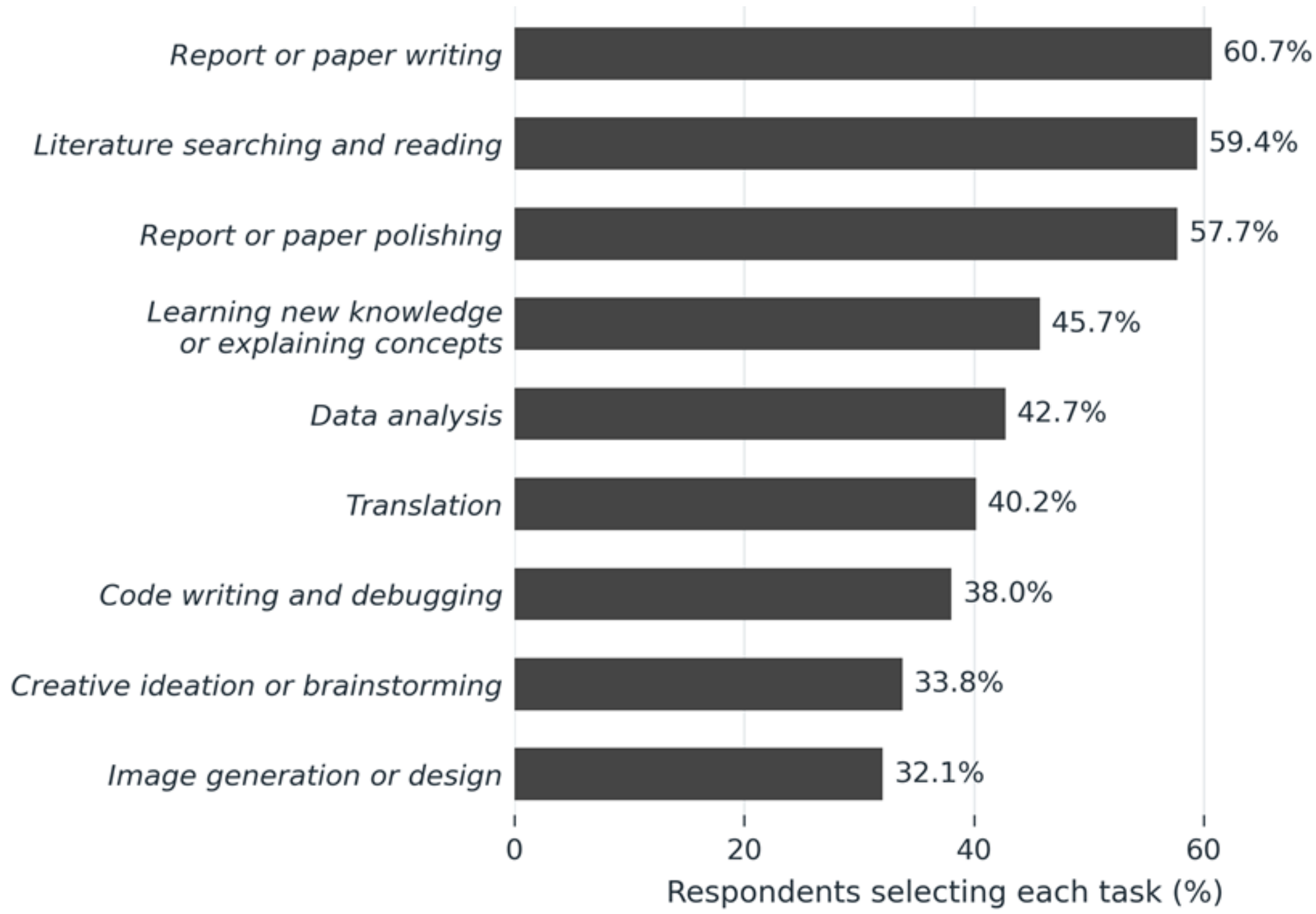

Figure 1. Academic tasks supported by GenAI tools. Multiple selections were allowed.

### 4.3 Orientations toward AI technologies

Figure 2 showed generally positive orientations toward AI technologies. Within affective orientation, interest in learning about AI (67.5%) and eagerness to explore new AI technologies (67.1%) were endorsed more frequently than confidence in completing AI-related tasks (62.0%). Within behavioral intention, continued use received the highest endorsement (75.7%), followed by exploring new application features (67.6%) and discussing AI-related topics with peers (64.1%). These results suggested that willingness to continue using AI was more widespread than confidence or intentions involving more active exploration and peer engagement. The overall positive orientation was broadly consistent with favorable student perceptions of GenAI reported by Chan & Hu (2023).

For perceived cognition, 71.8% of respondents endorsed their ability to use AI to solve practical problems, 67.1% endorsed evaluating the suitability of AI applications, and 58.2% endorsed distinguishing relevant technical concepts. The conceptual-differentiation item also had the largest neutral response share (28.2%), suggesting greater uncertainty in this area than in practical application or evaluation. Ethical orientation was comparatively consistent across items: 69.3% recognized substantial risks associated with misuse, 71.8% supported adherence to ethical and legal standards, and 69.7% affirmed user responsibility. The combined emphasis on practical application, evaluation, and ethical responsibility was consistent with multidimensional accounts of AI literacy, in which understanding, application, evaluation, and responsible use are related but distinct components (Long & Magerko, 2020; Ng et al., 2021).

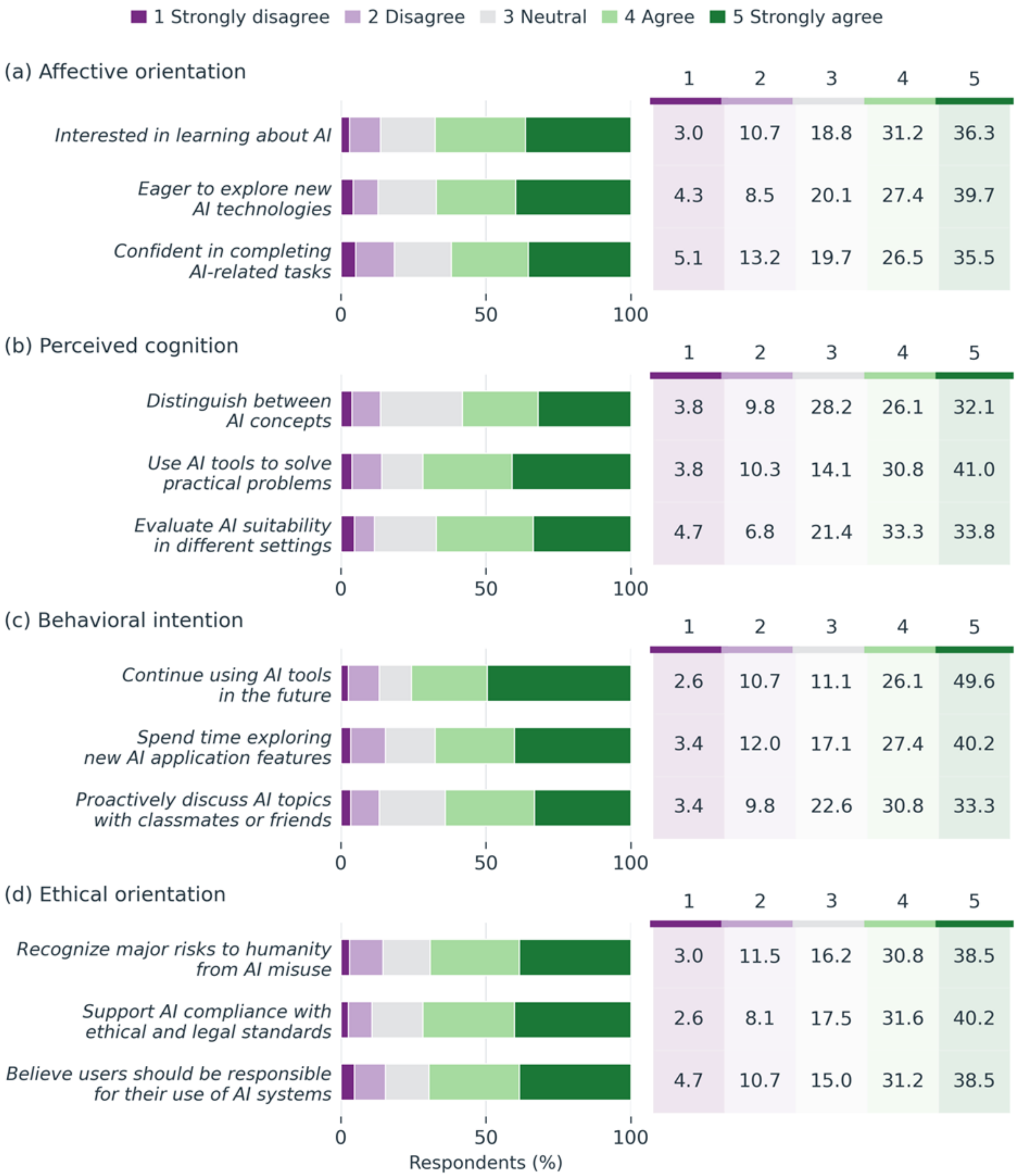

Figure 2. Response distributions for (a) affective orientation, (b) perceived cognition, (c) behavioral intention, and (d) ethical orientation. For each item, the stacked bar and five numerical columns present the same response distribution; this format is also used in subsequent figures of the same type.

### 4.4 Perceived prompting capability

As shown in Figure 3, endorsement with the five prompting-capability items ranged from 66.2% to 72.2%, indicating generally positive self-assessments. Including task-relevant details, such as background, scope, or intended use, was endorsed by 70.9% of respondents, while 67.1% reported confidence in breaking complex tasks into steps.

Prompt refinement after reviewing an initial response received the highest endorsement (72.2%), and 70.5% reported confidence in modifying prompts when outputs contained factual errors or bias. Together, these results suggest that iterative adjustment was a relatively strong aspect of respondents' perceived prompting capability and that prompting was commonly viewed as an interactive process rather than a one-off instruction. This pattern is consistent with Aruleba et al. (2025), who likewise emphasized refinement and appraisal of generated responses in students' prompting practices.

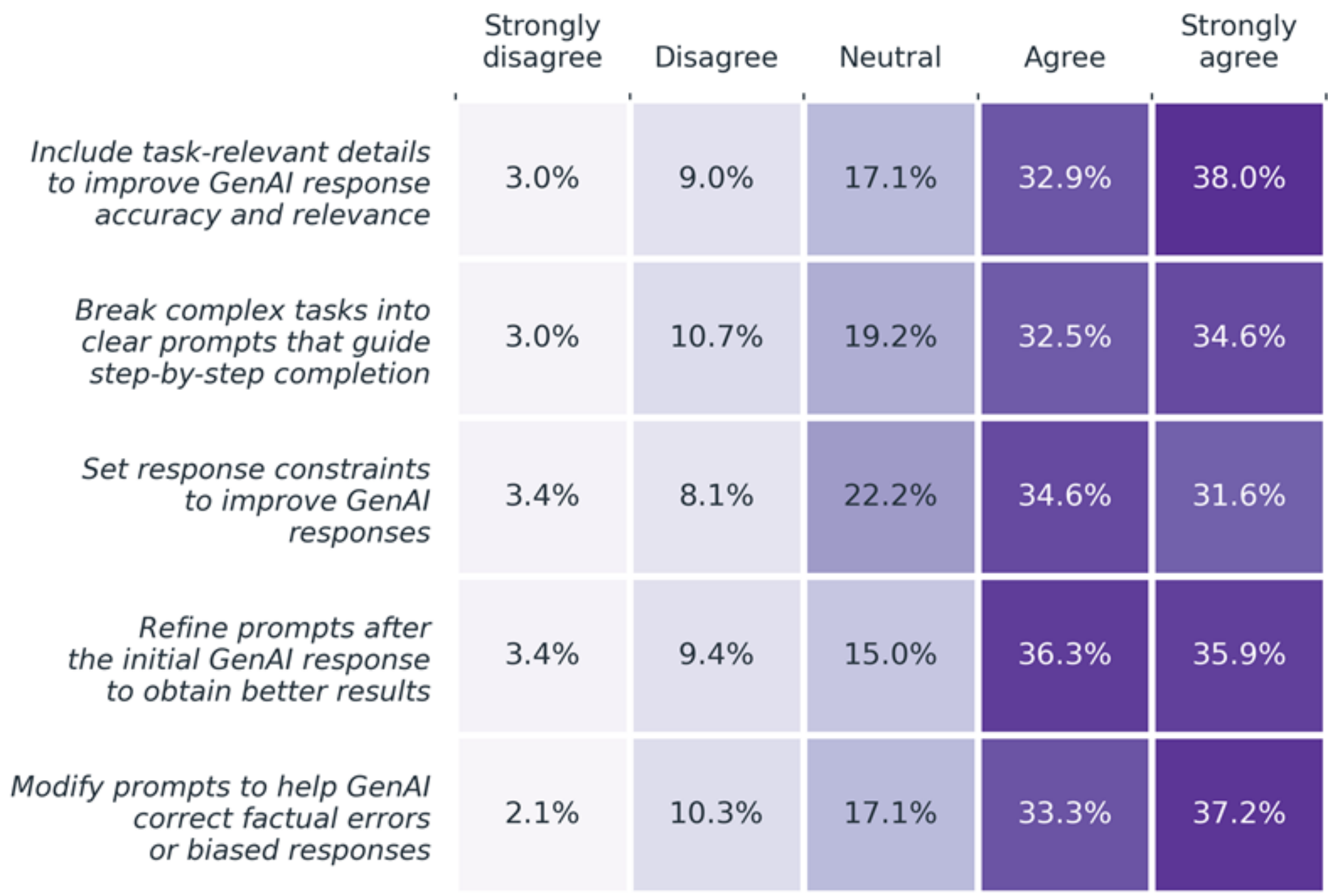

Figure 3. Self-reported GenAI prompting capability in academic tasks

Setting response constraints, such as role, tone, length, or format, received the lowest agreement (66.2%) and the largest neutral share (22.2%). Relative to iterative refinement and provision of task details, respondents therefore appeared somewhat less certain about their ability to specify and combine output constraints. The distributions suggested comparatively greater uncertainty about explicit output control than about refinement or contextualization. However, because these measures captured self-perceived capability rather than directly observed performance, they should be interpreted as indicators of perceived readiness and confidence rather than demonstrated competence.

### 4.5 Prompting strategies across task contexts

In the research-oriented assignment scenario, respondents considered using GenAI to prepare a course assignment on the impact of GenAI on university students' employment. As shown in Figure 4(a), revising the prompt or adding information after an unsatisfactory response received the highest agreement (74.3%), followed by providing relevant background before asking the main question (68.8%). Requesting a specific output format and step-by-step analysis each received 59.4% agreement, while assigning a topic-relevant research-expert role received 58.2%. Thus, iterative revision and contextual information were more strongly endorsed than more explicitly structured techniques such as role assignment and output formatting.

In the creative workshop preparation scenario, respondents considered developing an appealing title and slogan for a workshop on "GenAI and the Future of Education". As shown in Figure 4(b), providing event background received the highest agreement (75.6%), followed by revising prompt requirements after an unsatisfactory response (72.7%), specifying the output format (72.2%), stating a desired creative direction or evaluation criteria (70.6%), and assigning an appropriate role (65.8%). Compared with the research-oriented scenario, endorsement was generally higher and more consistent across strategies.

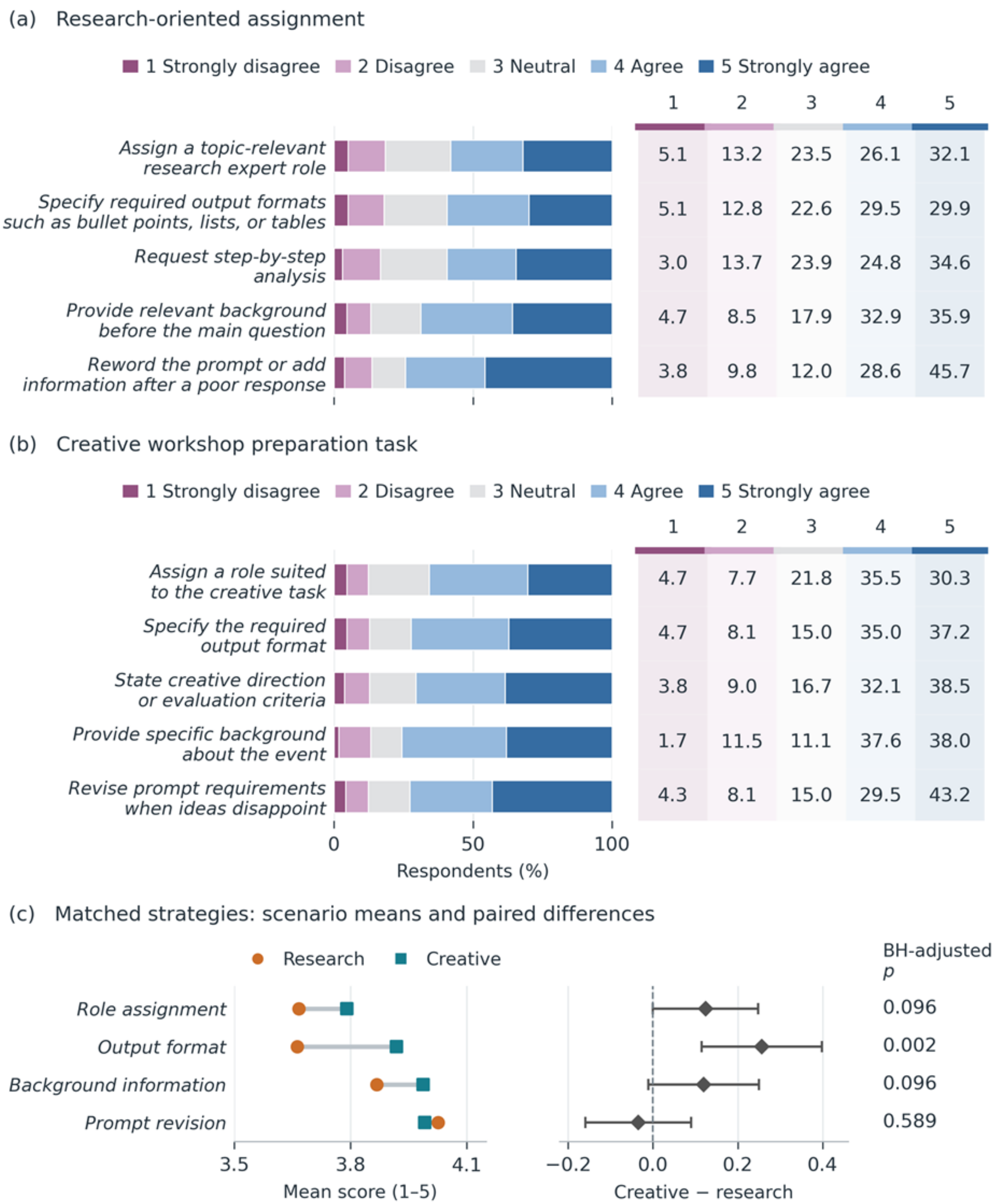

Figure 4. Scenario-specific GenAI prompting strategies: (a) research-oriented assignment scenario; (b) creative workshop preparation scenario; and (c) scenario means (left) and Creative-minus-Research differences with pointwise 95% confidence intervals (right). *p* values are BH-adjusted across the four paired comparisons.

Across both scenarios, providing background information and revising prompts were among the most strongly endorsed strategies, suggesting that respondents viewed contextualization and iterative adjustment as broadly applicable across tasks. From a self-regulated learning perspective, iterative revision may reflect monitoring and adjustment in response to task feedback (Panadero, 2017), although the present self-report data did not capture the quality of such monitoring.

Figure 4(c) and Table 3 present the four conceptually matched strategy comparisons. The third item in each scenario was excluded because the two items represented substantively different strategies. After BH adjustment, only output-format specification differed significantly between scenarios. This strategy was reported more strongly in the creative scenario, with a mean difference of 0.26 points on the five-point scale, $dz = 0.233$, and BH-adjusted $p = 0.002$. Differences in role assignment, background information, and iterative revision were not statistically significant after adjustment.

The greater emphasis on output-format specification in the creative scenario likely reflected the task's clearly defined deliverable (i.e., a title and slogan), which made explicit formatting requirements more salient. In contrast, background information and iterative revision showed no significant contextual differences after adjustment, suggesting that respondents regarded these strategies as broadly applicable across both tasks. Overall, the findings indicate that prompting strategies were generally consistent across contexts, with selective adaptation where specific task demands made particular strategies more relevant. However, these results reflect reported preferences in hypothetical scenarios rather than observed prompting behavior in authentic tasks.

Table 3 Matched scenario comparisons. Research denotes the research-oriented assignment scenario; Creative denotes the workshop-title and slogan creation scenario.

| **Strategy** | **Research *M*** | **Creative *M*** | **Difference [95% confidence interval]** | $d_z$ | ***p*** | **BH-adjusted *p*** |
|---|---|---|---|---|---|---|
| Role assignment | 3.67 | 3.79 | 0.12 [0.00, 0.25] | 0.129 | 0.050 | 0.096 |
| Output format | 3.66 | 3.92 | 0.26 [0.11, 0.40] | 0.233 | < 0.001 | 0.002 |
| Background data | 3.87 | 3.99 | 0.12 [-0.01, 0.25] | 0.118 | 0.072 | 0.096 |
| Iterative revision | 4.03 | 3.99 | -0.03 [-0.16, 0.09] | -0.035 | 0.589 | 0.589 |

### 4.6 General prompting strategies

As shown in Figure 5, endorsement was highest for organizing longer prompts using line breaks, paragraphs, or punctuation (70.6%), followed by providing relevant details at the outset rather than beginning with a short trial instruction (69.2%). Brief and direct instructions were endorsed by 63.7% of respondents, while 59.4% endorsed using a fixed template, such as role–task–output format. The template item also had the largest neutral share (24.4%), suggesting less consistent preference for standardized prompt structures.

The results indicate that conciseness and completeness were not necessarily viewed as competing strategies. Respondents commonly endorsed both brief instructions and the provision of relevant information from the outset, suggesting that effective prompting may depend more on selecting and organizing task-relevant information than on prompt length alone. This interpretation is broadly consistent with Lo's (2023) CLEAR framework, which emphasizes conciseness alongside explicitness and iterative refinement.

The relatively strong endorsement of organizing longer prompts further suggests that respondents valued structural clarity when more information was needed. In contrast, the lower endorsement of fixed templates points to greater variation in preferences for standardized prompting formats. Overall, the findings indicate diverse prompting preferences rather than a single dominant style.

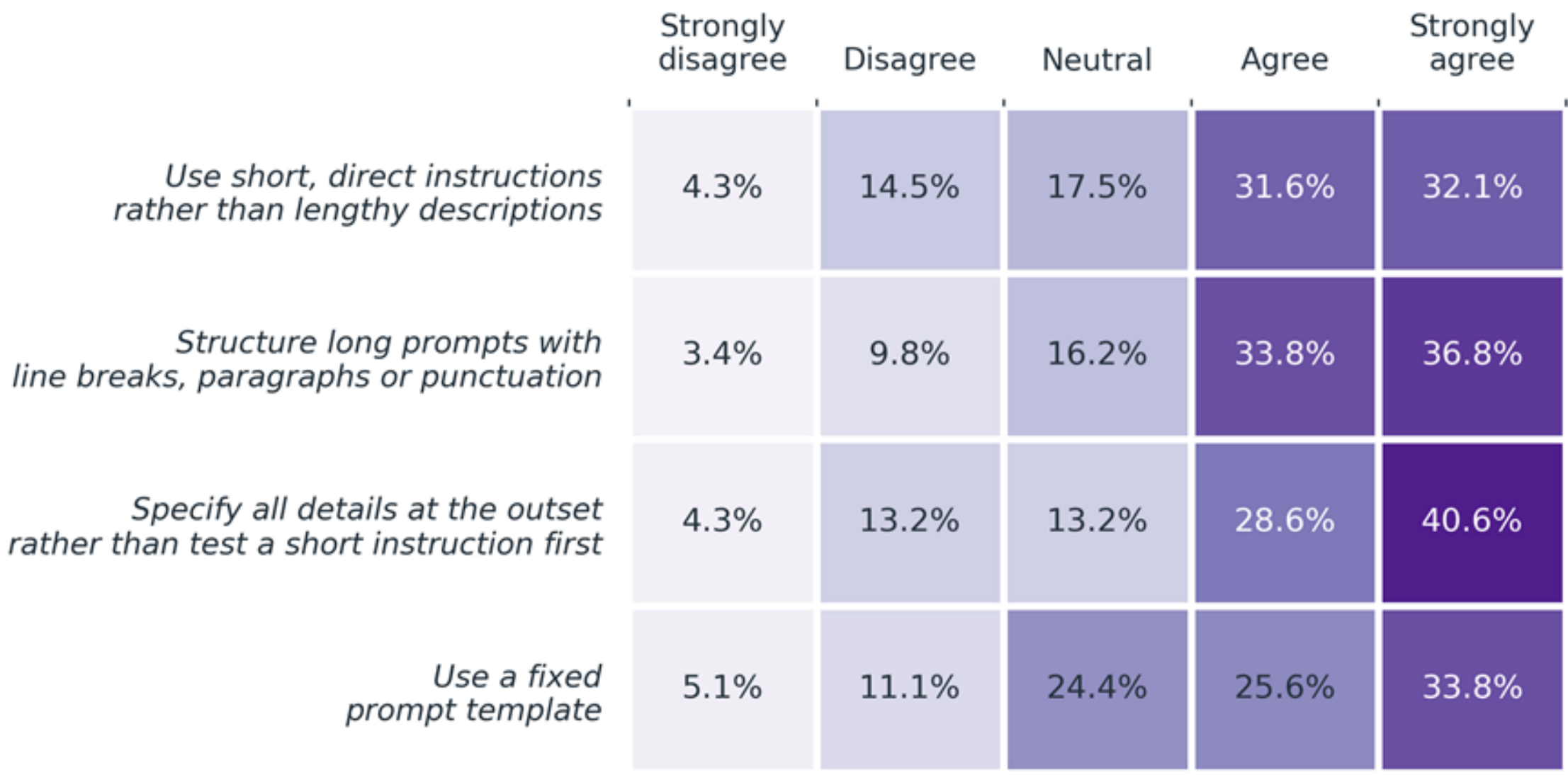

Figure 5. General prompting strategies in academic tasks

### 4.7. Reported prompting difficulties

Substantive responses to the optional open-ended question were provided by 12.8% of respondents. Among those who commented, the most common themes were prompt–intent alignment (40.0%), prompt-formulation difficulty (23.3%), output usefulness or presentation (20.0%), iteration or conversational context (16.7%), and factual verification (13.3%), with smaller frequencies across other miscellaneous categories. Because multiple codes could be assigned to each response and only a minority of respondents provided comments, these percentages describe the commenting subset rather than the prevalence of difficulties in the full sample.

Several comments illustrated these challenges. One respondent noted, "Even though I have described things in great detail, it still does not understand what I mean", while another reported "difficulty defining specialized problems precisely". A third stated, "After repeated prompts requesting revisions, the GenAI becomes confused, leading to poorer output quality". These comments suggest that adding detail or repeatedly revising a prompt may not resolve an interaction when the task objective, relationships among requirements, or criteria for an acceptable response remain unclear.

Previous studies have similarly reported difficulties in expressing intent, refining prompts, and evaluating generated responses (Zamfirescu-Pereira et al., 2023; Nguyen et al., 2024; Aruleba et al., 2025). The present findings highlight prompt–intent alignment as a key challenge: students may be willing to revise prompts yet still struggle to translate their goals into instructions that GenAI can interpret accurately. University training should therefore emphasize clarifying task goals and expressing them through explicit requirements, relevant context, constraints, and criteria for acceptable outputs.

### 4.8. Measurement models and structural associations

Table 4 compared two measurement models for the same 27 rating items. In the one-factor model, all items loaded on a single latent dimension, whereas in the seven-correlated-factor model, items loaded on their questionnaire-defined domains (D1–D7) and the seven factors were allowed to correlate. The latter therefore represented the domains as related but distinct rather than as independent competencies.

The seven-correlated-factor model showed substantially better fit than the one-factor model, with $\chi^2$ = 538.95, $p < 0.001$, CFI = 0.956, TLI = 0.949, RMSEA = 0.058, and SRMR = 0.036, compared with CFI = 0.847, TLI = 0.835, RMSEA = 0.104, and SRMR = 0.053 for the one-factor model. Standardized loadings in the seven-factor model ranged from 0.733 to 0.905. These results supported retaining the questionnaire-defined multidimensional structure rather than reducing all 27 items to a single general factor.

However, the latent factors were highly correlated, indicating substantial overlap among the domains. Thus, the better fit of the seven-factor model supported multidimensionality but did not demonstrate complete separation between constructs. Some shared covariance may also have reflected the common self-report format (Podsakoff et al., 2003).

Table 4 Exploratory measurement model comparison

| **Model** | **$\chi^2$ (df)** | **CFI** | **TLI** | **RMSEA [90% confidence interval]** | **SRMR** |
|---|---|---|---|---|---|
| One general factor | 1137.25 (324) | 0.847 | 0.835 | 0.104 [0.097, 0.110] | 0.053 |
| Seven correlated factors | 538.95 (303) | 0.956 | 0.949 | 0.058 [0.050, 0.066] | 0.036 |

Table 5 Complete exploratory structural associations

| **Outcome** | **Predictor** | ***β* [95% confidence interval]** | ***p*** | **BH-adjusted *p*** |
|---|---|---|---|---|
| Perceived prompting capability | Affective orientation | 0.177 [-0.065, 0.419] | 0.152 | 0.265 |
| | Perceived cognition | 0.390 [0.161, 0.619] | < 0.001 | 0.012 |
| | Behavioral intention | 0.241 [0.028, 0.455] | 0.027 | 0.094 |
| | Ethical orientation | 0.159 [-0.042, 0.360] | 0.121 | 0.241 |
| Research-oriented assignment | Affective orientation | 0.068 [-0.217, 0.353] | 0.639 | 0.639 |
| | Perceived cognition | 0.195 [-0.096, 0.487] | 0.189 | 0.265 |
| | Behavioral intention | 0.138 [-0.116, 0.392] | 0.288 | 0.365 |
| | Ethical orientation | 0.105 [-0.099, 0.308] | 0.313 | 0.365 |
| | Perceived prompting capability | 0.427 [0.084, 0.769] | 0.015 | 0.094 |
| Creative workshop preparation | Affective orientation | 0.263 [0.004, 0.522] | 0.047 | 0.131 |
| | Perceived cognition | 0.160 [-0.075, 0.394] | 0.182 | 0.265 |
| | Behavioral intention | 0.073 [-0.160, 0.305] | 0.540 | 0.581 |
| | Ethical orientation | 0.220 [0.027, 0.413] | 0.025 | 0.094 |
| | Perceived prompting capability | 0.240 [-0.036, 0.515] | 0.088 | 0.205 |

Table 5 reports the 14 exploratory structural associations. Because multiple correlated predictors were entered simultaneously, each coefficient represented the unique association with the outcome after accounting for the other predictors in the same equation. Perceived cognition showed the strongest association with perceived prompting capability, $\beta = 0.390$, and was the only path that remained significant after BH adjustment, adjusted $p = 0.012$. Thus, higher perceived cognition was associated with greater perceived prompting capability after controlling for affective orientation, behavioral intention, and ethical orientation. This finding is consistent with AI literacy frameworks linking understanding and application with the effective and evaluative use of AI technologies (Long & Magerko, 2020; Ng et al., 2021).

Several other positive associations were evident before multiplicity adjustment but did not remain statistically significant. Behavioral intention was associated with perceived prompting capability ($\beta = 0.241$, $p = 0.027$), while perceived prompting capability was associated with research-oriented assignment strategies ($\beta = 0.427$, $p = 0.015$). In the creative workshop scenario, affective orientation ($\beta = 0.263$, $p = 0.047$) and ethical orientation ($\beta = 0.220$, $p = 0.025$) also showed positive unadjusted associations, but neither met the adjusted significance criterion.

These non-significant paths nevertheless had non-trivial point estimates. Most notably, the association between perceived prompting capability and research-oriented assignment strategies was relatively substantial, but its uncertainty and failure to survive multiplicity adjustment limited evidence for an independent effect. This pattern is plausible when correlated predictors are modeled simultaneously, as shared variance can increase uncertainty in individual coefficients and make their unique contributions difficult to distinguish (Dormann et al., 2013). Adjustment across 14 tests further raised the evidentiary threshold for individual paths (Benjamini & Hochberg, 1995). The small number of adjusted-significant associations should therefore be interpreted as limited evidence for unique conditional relationships, rather than as evidence that the domains were unrelated.

Since the data were cross-sectional and the structural specification was covariance-equivalent to the measurement model, the estimated paths should not be interpreted as directional or causal. Overall, the findings support viewing GenAI-related orientations, perceived prompting capability, and task-specific prompting strategies as closely related but conceptually distinct domains (Woo et al., 2026).

## 5 Discussion

### 5.1 From prompt refinement to evaluative self-regulation

Iterative refinement was among the most strongly endorsed aspects of both perceived prompting capability and scenario-specific strategy preferences, suggesting that students recognized the value of adjusting their interaction with GenAI after an initial response. However, the open-ended comments revealed an important distinction between revising a prompt and diagnosing what needs to be revised. Some respondents described providing extensive detail without obtaining the intended response, whereas others reported that repeated revisions made the interaction less coherent.

These comments point to prompt–intent alignment as an important challenge. Simply adding information or repeatedly modifying a prompt may not improve an interaction when the task goal, relationships among requirements, or criteria for an acceptable response remain unclear. Effective refinement therefore requires users first to identify the source of the mismatch and then decide which aspect of the prompt or interaction should be changed.

This process closely resembles self-regulated learning, particularly cycles of planning, monitoring, evaluation, and adaptation (Zimmerman, 2002; Panadero, 2017). In GenAI-supported learning, students may define the task, evaluate a generated response against explicit criteria, diagnose discrepancies, and make targeted revisions. Zamfirescu-Pereira et al. (2023) similarly showed that nonexpert users can struggle to evaluate and refine prompts systematically. The present findings therefore suggest that willingness to revise prompts should be distinguished from the ability to diagnose why an output does not align with the user's intention.

The structural findings reinforce this interpretation. Perceived cognition was the only orientation domain that retained a statistically supported unique association with perceived prompting capability after correction for multiple testing. Because this domain encompassed understanding AI concepts, applying AI tools to practical problems, and judging their suitability, the result suggests that perceived prompting capability was more closely related to perceived understanding and evaluative application than to positive attitudes or intentions alone. This interpretation is consistent with multidimensional accounts of AI literacy that integrate understanding, application, and critical evaluation (Long & Magerko, 2020; Ng et al., 2021). Nevertheless, because both constructs were self-reported and conceptually related, the association should not be interpreted as directional or causal.

### 5.2 Pedagogical implications

The scenario findings indicate that prompting competence combines transferable principles with task-sensitive adaptation. Providing relevant background and revising an initial prompt were among the most strongly endorsed strategies across both scenarios, suggesting that contextualization and iterative refinement have broad applicability. By contrast, output-format specification was more strongly endorsed for the creative task, where the nature of the required deliverable made format particularly salient. This pattern supports the view of prompting as a context-sensitive practice in which general strategies are adapted to task goals and expected outputs (White et al., 2023; Aruleba et al., 2025; Woo et al., 2026). Training could therefore introduce broadly useful principles while helping students determine when additional specifications concerning format, role, constraints, or evaluation criteria are warranted. The comparatively lower endorsement of fixed prompt templates reinforces this interpretation: templates may provide useful initial scaffolding, but they should not be treated as universally applicable solutions.

Difficulties with prompt–intent alignment also point to a role for assessment in encouraging more deliberate prompting practices. This implication is consistent with research distinguishing successful GenAI-assisted products from the development of learning, transfer, and evaluative judgment (Fan, Tang, et al., 2025). Where GenAI use is permitted, students could be asked to present selected prompts or revisions and briefly explain the reasoning behind important changes. The purpose would not be to reward elaborate or formulaic prompting, but to make visible how students interpreted the task, evaluated generated responses, and adapted their interaction. Such an approach is also consistent with constructive alignment, because evidence of prompting practice would remain tied to the disciplinary learning outcomes that the assessment is intended to develop (Biggs, 1996).

The association analyses provide further support for connecting prompting with broader AI literacy. Perceived cognition was the only orientation domain that retained a significant unique association with perceived prompting capability after adjustment for multiple testing. Because this domain encompassed students' perceived ability to apply AI to practical problems, evaluate the suitability of AI applications, and distinguish relevant concepts, the finding suggests that prompting capability is more closely related to understanding and evaluative use than to positive attitudes toward AI alone. The association is correlational and should therefore not be interpreted causally, but it nevertheless reinforces the educational importance of developing students' capacity to make informed judgments about GenAI use. This is particularly relevant given students' relatively strong intention to continue using AI: an important institutional priority is not simply to facilitate access or encourage adoption, but to help students judge when, why, and how GenAI can be used appropriately in academic work.

These findings suggest that prompting instruction should be embedded within broader AI literacy rather than taught primarily as a set of technical techniques. Effective interaction with GenAI involves not only formulating instructions, but also articulating intentions, judging whether GenAI is appropriate for a task, identifying the information it requires, evaluating generated outputs, and revising prompts when necessary (Nguyen et al., 2024; Zamfirescu-Pereira et al., 2023). Prompting education should therefore develop evaluative judgment alongside operational skill, helping students interpret tasks, define criteria for acceptable responses, assess output quality, and justify revisions. Framed in this way, prompting becomes part of the wider processes of critical judgment and adaptive learning that higher education seeks to develop.

### 5.3 Limitations and future research

Several limitations constrain interpretation of the findings. First, the cross-sectional questionnaire measured perceived capability and reported preferences rather than directly observed prompting performance. Agreement with a strategy therefore did not establish how frequently or effectively students used it, and the two scenarios necessarily simplified authentic academic work.

Second, all principal domains used self-report measures collected within the same questionnaire. Their substantial correlations may partly reflect shared response tendencies or common-method effects (Podsakoff et al., 2003). Conceptual proximity between perceived cognition and perceived prompting capability may also have contributed to their association. Future scale development should further clarify construct boundaries through content validation, cognitive interviewing, and more differentiated item wording (Boateng et al., 2018).

Third, the optional comments were provided by only a small subset of participants and were illustrative rather than representative. Prompt–intent alignment should therefore be regarded as an important qualitative indication for further investigation rather than as the predominant difficulty in the full sample.

Future studies should combine self-report measures with authentic prompting tasks in which prompts, revisions, outputs, verification behavior, and final decisions can be examined directly. Longitudinal or intervention research could determine whether students improve in aligning prompts with intended goals, selecting task-appropriate strategies, diagnosing inadequate outputs, and evaluating generated information. Such work would help distinguish perceived prompting capability from demonstrated competence and clarify whether improved prompting contributes to disciplinary learning and transfer.

Overall, the findings suggest that prompting education should extend beyond fixed formulas. Its central educational value may lie in helping students translate academic intentions into task-appropriate prompts and diagnose mismatches between intended and generated outcomes. In this sense, prompting can provide a context for developing academic judgment and self-regulation rather than merely a means of obtaining better GenAI outputs.

## 6 Conclusion

This study examined students' perceived GenAI prompting capability and reported prompting strategies. Students generally expressed confidence in providing relevant context and refining prompts after an initial response, whereas specifying response constraints received somewhat weaker endorsement. These findings suggest that contextualization and iterative refinement were prominent aspects of students' self-perceived prompting capability, while more explicit control of output requirements appeared less consistently established.

The two scenarios further indicated that reported prompting strategies were partly sensitive to task context. Providing background information and revising prompts were strongly endorsed across both tasks, whereas output-format specification was more strongly endorsed in the creative workshop scenario and was the only matched strategy that differed significantly after multiplicity adjustment. This pattern suggests that some strategies may be broadly applicable across tasks, while others are adapted to particular task demands. Similarly, the endorsement of both concise instructions and detailed upfront information suggests that students did not necessarily view brevity and completeness as competing approaches; rather, their preferences appeared to depend on how task-relevant information was selected and organized.

Perceived cognition showed the clearest unique association with perceived prompting capability, linking higher self-assessed understanding, practical application, and evaluation of AI suitability with greater perceived prompting capability. However, the strong correlations among the domains and uncertainty surrounding most other structural paths caution against interpreting these relationships as distinct or causal influences. Instead, AI-related orientations, perceived prompting capability, and task-specific strategy preferences appear to represent closely related but non-interchangeable aspects of students' engagement with GenAI.

The optional comments, although provided by only a minority of respondents, highlighted prompt–intent alignment as a key challenge. Some students appeared willing to revise prompts but still struggled to translate their intended goals into instructions that GenAI could interpret appropriately. University training should therefore move beyond fixed prompt templates and focus on helping students clarify task goals, specify relevant context and constraints, and diagnose misalignment when outputs do not meet the intended purpose.

**Data availability**

Data will be made available on reasonable request to the corresponding author.

**Competing interest**

The authors declare no competing interests.

**Ethics approval**

The study protocol was approved by the University Research Ethics Review Panel of Xi'an Jiaotong-Liverpool University (XJTLU) on 8 Jun 2026. All procedures were conducted in accordance with applicable ethical guidelines and regulations required by XJTLU Policy on Ethical Conduct in Research.

**Informed consent**

Informed consent was obtained from all individual participants for participation in this study. Anonymity of their identification and voluntary participation were assured. No identifiable data of participants were present within the manuscript.